# Scalable and Efficient Self-Join Processing technique in RDF data

Awny Sayed [1] and Amal Almaqrashi [2]

[1] Faculty of Science, Minia University, Egypt
Ibri College of Applied Sciences, Sultanate of Oman
awny.ibr@cas.edu.om

[2] Ibri College of Applied Sciences, Sultanate of Oman
amalsultan.ibr@cas.edu.om

**Abstract**
Efficient management of RDF data plays an important role in successfully understanding and fast querying data. Although the current approaches of indexing in RDF Triples such as property tables and vertically partitioned solved many issues; however, they still suffer from the performance in the complex self-join queries and insert data in the same table. As an improvement in this paper, we propose an alternative solution to facilitate flexibility and efficiency in that queries and try to reach to the optimal solution to decrease the self-joins as much as possible, this solution based on the idea of "*Recursive Mapping of Twin Tables*". Our main goal of *Recursive Mapping of Twin Tables (RMTT) approach* is divided the main RDF Triple into two tables which have the same structure of RDF Triple and insert the RDF data recursively. Our experimental results compared the performance of join queries in vertically partitioned approach and the RMTT approach using very large RDF data, like DBLP and DBpedia datasets. Our experimental results with a number of complex submitted queries shows that our approach is highly scalable compared with RDF-3X approach and RMTT reduces the number of self-joins especially in complex queries 3-4 times than RDF-3X approach.
**Keywords:** *Semantic Web, RDF, RDF Storing, RDF Indexing, Self-join, Query Processing.*

## 1. Introduction

There are several initiatives to improve the situation and reducing the drawbacks of the current web. One of them is a *Semantic Web*, which is coined by the W3C founder Tim Berners-Lee in a Scientific American article that is describing the future of the Web [1]. The Semantic Web would give more structure and computer-understandable meaning as well as provide a common framework for data sharing across applications, enterprises, and communities. The core of the Semantic Web is built on the *Resource Description Framework (RDF)* data model [2] [3] [4]. An RDF store consists of a collection of statements, called triples, of the form *(subject, predicate, and object)* also known as *(subject, property, and value)*, where subject and predicate are resource URIs and O is either a URI or a literal value.

RDF storage has witnessed numerous research initiatives in varied domains. Despite the best efforts, a scalable, efficient and fast index has eluded researcher's attention. A typical RDF data-store consist of billions of triples (a triple comprise Subject, Predicate & Object) with extensive and wide range of self- dependencies among the subject and the object field values. These results in recursive self joins with an added cost to the query optimizer [1]. Besides self-joins, unions and null values also create far greater performance related issues. There exist broadly two ways to deal with these issues: either to re-design the RDF data-store from scratch using a new setup for representing the triples along with the modified query engine design or to explore faster and more efficient indexing strategies that provide impeccable query processing times irrespective of scalability. This paper focuses on the second aspect related to a new index design.

The rest of the paper is organized as follows. Section 2 introduces the research efforts approaches to store and query RDF data efficiently. Section 3 provides the details of our proposed solution (RMTT) for storing and querying RDF data. Finally, section 4 concludes the paper and provides some suggestions to improve the research direction on our methodology as well as reducing join problems.

## 2. Related Work

In this section, we discuss the state of storing RDF data in Triple table, with an extended look at the property table approach and vertically partitioned approach.
The RDF data stores in a single table which consists of three columns (S, P, O), and each triple has subject, predicate, and object respectively. This approach is called *triples table approach* [5]. The serious performance issue of this approach is all the triples stored in a single RDF table which requires expensive and complex self joins over the triples table as pointed out in [11, 12, and 13]. Thus, as queries become more complex the execution and time increase. In addition, it is exceeding the memory size and congestion of the RDF data sets. Nevertheless, this approach has been implemented by systems like Oracle [14], 3store [5], Redland [20], RDFStore [21] and rdfDB [22]. The research community later introduces an alternative solution for improving the triples table and minimize the number of self-joins issues. An alternative methodology to the previous is the *property table approach* [6].

The property approach deformalized the RDF table that stored in a flattened format. Furthermore, it is classified into two types which are *property class table* and *clustered property table*. The clustered table contains clustered of properties that tend to be defined together. Whereas the property class table exploits the type property of subjects to cluster similar sets if subject together in the same table [7]. The most important advantage for representing the property tables is that they can reduce subject-subject self joins of the triples table. However, this approach may not fit well the RDF data because of unstructured data and missing properties. In an interpretation, not all properties will be defined for all subjects and that perhaps led to many NULLs value which increases the overhead in the memory space. Another problem with the property table is the abundance of multi-valued attributes found in RDF data which cause further complexity and with combine data from several tables the issue of improving the performance of self-joins queries maybe become poor. In summary, property tables
are rarely used due to their complexity and inability to handle multi-valued attributes. However, this approach has been used by tools like Sesame [23], Jena2 [12], RDFSuite [24] and 4store [25].

Abadi et al. [7] proposed *vertically partitioned approach* as an alternative solution to the property table to speed up queries and minimize its limitations and using a fully decomposed storage model (DSM) [10]. In this approach, an RDF table is rewritten into *n two-column* tables, where *n* is the number of unique properties. Moreover, the first column is *subject* and the second is an *object*.

One of the primary benefits of vertical partitioning is the support for rapid *subject-subject* joins. This feature is achieved by sorting the tables via subject as we mentioned above each binary table has subject and object columns. The tables being sorted by subject, one has a way to use fast merge joins to reconstruct information about multiple properties for subsets of subjects.

The experiments in these papers [8] and [9] showed that the vertical partitioning approach also performs poorly for querying RDF data and slow insertion, because of the multi property tables. In a spite, the vertical partitioning approach supports multi valued attributes and heterogeneous records. In addition, it is eliminating the subjects that don't define a particular property. Clearly, it reduces the NULLs value through that elimination.

Over the past decade, a fair number of RDF storage systems has been developed and implement the previous approaches, including Sesame[23], YARS [26], Kowari system[27], Virtuoso [28], RDF-3X [29], Hexastore [13], BitMat [30] etc. The following gives a simple overview on some of the previously mentioned systems.

*YARS system* [26] (Yet Another RDF Store) combines methods from Information Retrieval and Databases to allow for better query answering performance over RDF data. It stores RDF data persistently by using six B+ tree indices. It not only stores the subject, the predicate and the object, but also the context information about the origin of the data. To speed up keyword queries, the lexicon keeps an inverted index on string literals to allow fast full-text searches. In each B+ tree, the key is a concatenation of the subject, predicate, object and context. The six indices constructed cover all the possible access patterns of quads in the form (s, p, o, c) where c is the context of the triple (s, p, o). This representation allows fast retrieval of all triple access patterns.

*RDF-3X* [29] is an RDF storage system with advanced indexes and query optimization that eliminates the need of physical database design by the use of exhaustive indexes for all permutations of subject-property-object triples. Neumann *et al.* use a potentially huge triples table, with their own storage implementation underneath (as opposed to using an
RDBMS). They overcome the problem of expensive self-joins by creating a suitable set of indexes. All the triples are stored in a compressed clustered B+ tree.

*Hexastore* [13] takes also a similar approach to YARS. The framework is based on the idea of main-memory indexing of RDF data in a multiple-index framework. The RDF data is indexed in six possible ways, one for each possible ordering of the three RDF elements by individual columns. The representation is based on any order of significance of RDF resources and properties and can be seen as a combination of vertical partitioning [7] and multiple indexing approaches [31].

As a consequence, the current approaches of RDF stores that we discuss above triple table, property table approach and vertically partitioned approach, they still suffer from weak data locality which affects the storage and query performance costs. As an improvement of this paper which led to a stronger data locality, we proposed the Recursive Mapping of Twin Tables (RMTT) approach.

## 3. RMTT Approach

Recursive Mapping of Twin Tables (RMTT) is an approach to partition the RDF Triple which includes <subject, predicate, and object> triples into two tables that have the same structure of RDF Triple. In addition, this algorithm depends on the following idea whereas the subjects of the second table are subset of the objects in the first table; however, the vice versa isn't. In contrast, the subjects in the first table are not subset of the objects in the second table and vice versa. So, $s2 \subset o1$ and $o2 \not\subset s1$ whereas (s1 and o1 are subject and object of the first table) (s2 and o2 are subject and object of the second table). In contrast, $s1 \not\subset o2$ and $o2 \not\subset s1$. Moreover, the insertion of data depends on two direction (s1 $\rightarrow$ o2) AND (s2 $\rightarrow$ o1).

### 3.1 Pseudocode

The following figure explains the pseudo code that used to create our RMTT approach.

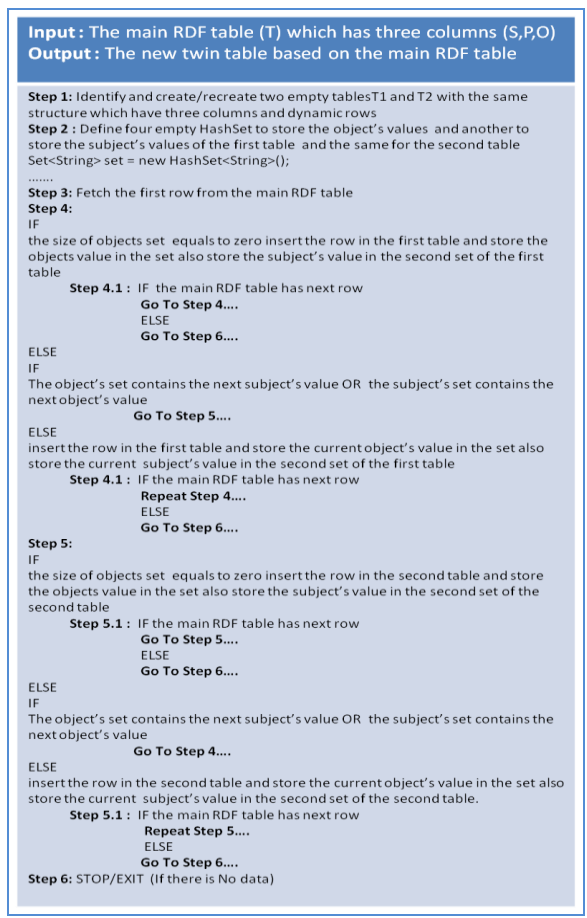

Figure 1: RMTT algorithm

This pseudocode describes how the algorithm will be going and divides the main RDF Triple into two table's recursively. Firstly, it defines two empty tables which have the same structure of the main RDF table. In addition, identify two empty sets for the first table to store the subject's values and object's values and the same for the second table. Then, fetch the first triple from the main and apply the RMTT algorithm. Note that, in this approach, inserting data done through two conditions which are the next subject's value in the first table is exist in all the previous object values or not , or the second condition is the next object's value in the first table is exist in all the previous subject values or not . If it already exists go to the next table and insert data and follow the same steps above. However , , insert data in the next triple in the same table and repeat the previous steps until the main triple has no data. The following DBLP example applies the RMTT algorithm which we proposed and gives you the final structure of the twin table as it shows in Figure 4.

### 3.2 Example

An example describes a magazine which publishes two articles. Each article has a title, year of publishing and it is written by one or more authors. An author has a name and some personal details which led to another branch such as affiliation or works in any university that has a name and place. For instance, Figure 2 is an example about al-Quds magazine published articles B1 and B2. Whereas B1 has a title Data Web which published in 2007 by Tom Lara. The author works at a University of Malta that is in Malta city. Figure 2 describes more details about this example. We must keep in mind that the RDF stores are increasing dynamically.

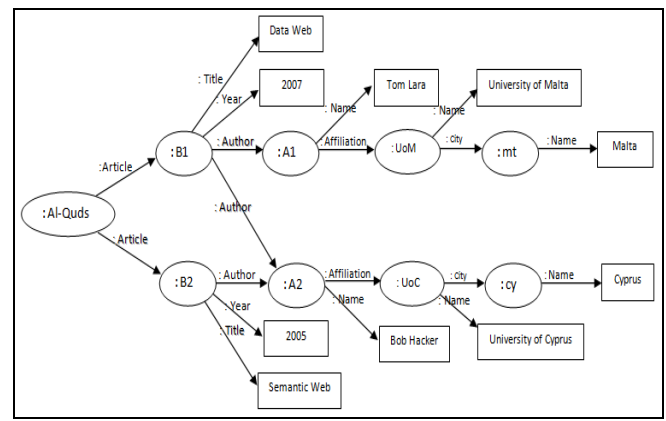

Figure 2: The graph model of DBLP example

| S | P | O |
|---|---|---|
| :B1 | rdf:Type | :Article |
| :B1 | :Title | "Data Web" |
| :B1 | :Year | 2007 |
| :B2 | rdf:Type | :Article |
| :B2 | :Title | "Semantic Web" |
| :B2 | :Year | 2005 |
| :B1 | :Author | :A1 |
| :B1 | :Author | :A2 |
| :B2 | :Author | :A2 |
| :A1 | rdf:Type | :Person |
| :A1 | :Name | "Tom Lara" |
| :A1 | :Affiliation | :UoM |
| :A2 | rdf:Type | :Person |
| :A2 | :Name | "Bob Hacker" |
| :A2 | :Affiliation | :UoC |
| :UoM | :Type | :University |
| :UoM | :Name | University of Malta |
| :UoM | :City | :mt |
| :mt | :Type | :City |
| :mt | :Name | "Malta" |
| :UoC | :Type | :University |
| :UoC | :Name | University of Cyprus |
| :UoC | :City | :cy |
| :cy | :Type | :City |
| :cy | :Name | "Cyprus" |

Figure 3: The main RDF table (MagazineTable)

"Table1"

| S | P | O |
|---|---|---|
| :A1 | rdf:Type | :Person |
| :A1 | :Name | "Tom Lara" |
| :A1 | :Affiliation | :UoM |
| :A2 | rdf:Type | :Person |
| :A2 | :Name | "Bob Hacker" |
| :A2 | :Affiliation | :UoC |
| :mt | :Type | :City |
| :mt | :Name | "Malta" |
| :cy | :Type | :City |
| :cy | :Name | "Cyprus" |

"Table2"

| S | P | O |
|---|---|---|
| :B1 | rdf:Type | :Article |
| :B1 | :Title | "Data Web" |
| :B1 | :Year | 2007 |
| :B2 | rdf:Type | :Article |
| :B2 | :Title | "Semantic Web" |
| :B2 | :Year | 2005 |
| :B1 | :Author | :A1 |
| :B1 | :Author | :A2 |
| :B2 | :Author | :A2 |
| :UoM | :Type | :University |
| :UoM | :Name | University of Malta |
| :UoM | :City | :mt |
| :UoC | :Type | :University |
| :UoC | :Name | University of Cyprus |
| :UoC | :City | :cy |

Figure 4 : the Twin table of RDF Data

Figure 3 describes the main RDF triple as we discuss its structure at the beginning of the paper which has the problem of complex self-joins in a single table. We will applies the RMTT algorithm on that table as follows. First of all, it fetches the data from the main table and store it in the array the process done triple by triple. Next, start to compare the next subject's value of the first table with all previous object's values in the same table if no similarity complete insertion and fetch the next triple from the main RDF data until find the difference. See figure 3 triple number 10 where the row's form as ( :A1, rdf:Type, :Person ). We can see the subject's value (: A1) is already exit with the previous object's values see figure 4 triple number 7 <:B1, :Author, :A>. Move to the second table and check the condition as we did with the previous table. Triple 12 in figure 3 allows moving to the neighbor or second table recursively and applying the same conditions until there is no unread data or entering to perform the algorithm.

### 3.3 Query Analysis

Query performance can be analyzed to improve it by viewing query execution plans or by manipulating of the queries .The goal of query performance is to minimize the response time of queries and to make the best use of the server's resources by minimizing the network traffic, disk I/O, and the efficiency of CPU time. The following tests queries from different levels to demonstrate the evaluation. These queries evaluate on the main RDF table as well as in the new twin tables as figures 3 and 4 were shown.

#### 3.3.1 Queries

Query (1): This is a simple triple query. It aims to retrieve the title of the article B1. With consider the table MagazineTable (S,P,O) which refers to the main RDF table as it is shown in figure 3.
SELECT T.O
FROM MagazineTable   T
WHERE T.S= ':B1'
AND T.P = ' :Title' ;
The output of execution this query is Data Web.

Query (2): This is a path query with one join. It aims to find all the authors of the article B2.

-------------------In Main RDF Table ------------------------------

SELECT T2.O   FROM MagazineTable T1 , MagazineTable T2
WHERE T1.O=T2.S AND T2.P=':Name' AND T1.S =':B2' AND T1.P=':Author';
The output of execution this query is Bob Hacker.

-------------------In Twin Table ------------------------------

SELECT  T2.O  FROM TABLE2 T1 , TABLE1 T2
WHERE T1.O=T2.S AND T2.P=':Name' AND T1.S =':B2' AND T1.P=':Author;
The output of execution this query is Bob Hacker.

Query (3): This is a path query with two joins. It aims to list all the articles which have authors from University of Malta and their authors.

-------------------In Main RDF Table ------------------------------

SELECT T3.S, T3.O FROM MagazineTable  T1 , MagazineTable T2 , RDF_TRIPLE  T3
WHERE T1.S = T2.O  AND T2.S = T3.O  AND T3.P=':Author' AND T2.P=':Affiliation' AND
T1.P=':Name' AND T1.O = 'University of Malta';
The output of execution this query is B1 and the author is A1.

-------------------In Twin Table ------------------------------

SELECT T1.S,T1.O FROM TABLE2 T1 , TABLE1 T2
WHERE (T1.S=T2.O OR T2.S=T1.O) AND ( T1.P='author') AND (T2.P='affiliation') AND T2.O = (select T2.S from TABLE2 T2 where T2.O = 'University of Malta')
 The output of execution this query is B1 and the author is A1.

Query (4): This is a star query which has four joins. It aims to display a list of authors who work on Cyprus and their articles.

-------------------In Main RDF Table ------------------------------

SELECT  T4.S, T4.O FROM MagazineTable T1 , MagazineTable T2 , MagazineTable T3,  MagazineTable T4 where T1.S = T2.O  AND T2.S = T3.O AND T3.S=T4.O AND T1.P=':Name' AND T1.O='Cyprus'
The output of execution this query is A2 , the articles B1 and B2.

--------------------In Twin Table -------------------------------

SELECT T2.S,T2.O from TABLE1 T1 , TABLE2 T2
where  (T2.S=T1.O OR T1.S=T2.O) AND ( T2.P=':Author' OR T2.P=':City') AND T1.O= (select T2.S from TABLE2 T2 where T2.O =(select T1.S from TABLE1 T1 where T1.O = 'Cyprus'))
The output of execution this query is A2 , the articles B1 and B2.

## 4. Experimental Results

This section studies the performance of RMTT approach on a heterogeneous experimental setup comprising real-world RDF datasets from different are of knowledge. We study query performance against RDF-3X triple (which based on one large triple table) and RMTT approach (which based on two small join tab tables), evaluating the number joins and the query response time. We compare our results with a highly-efficient store (RDF 3X).

4.1 Datasets

The datasets that we use in our experimental are: CAS-IBRI (holds information about staff, faculty, students, departments), DBLP (Provides information about computer science journals and proceedings), DBpedia (is the semantic evaluation of Wikipedia).  The main characteristics of these datasets are described in the following table:

| Dataset | Size(MB) | #Triples | #Subjects | #Predicates | #Objects |
|---|---|---|---|---|---|
| CAS-IBRI | 130.5 | 1,050,00 | 290,960 | 25 | 369,233 |
| DBLP | 7,639,60 | 111,234,500 | 7,534,606 | 29 | 18,234,612 |
| DBpedia | 29,423,60 | 218,345,10 | 17,423,222 | 33,456 | 63,654,706 |

Table 1: dataset descriptions

4.2 Query Performance

We submitted number of complex queries against RMTT approach and RDF-3X approach focusing on query response time.

| Queries | Times (second) | | | | | | No. of Self-joins (based DBedia) | |
|---|---|---|---|---|---|---|---|---|
| | CAS-Ibri | | DBLP | | DBpedia | | RMTT | RDF-3X |
| | RMTT | RDF-3X | RMTT | RDF-3X | RMTT | RDF-3X | | |
| Q1 | 0.0187 | 0.01 | 10.1439 | 40.3245 | 12.1437 | 36.2308 | 1 | 1 |
| Q2 | 0.2124 | 0.214 | 17.3453 | 47.4133 | 18.3455 | 36.0912 | 1 | 1 |
| Q3 | 0.1127 | 0.216 | 18.1436 | 50.3240 | 22.1438 | 60.4002 | 2 | 2 |
| Q4 | 0.3424 | 0.344 | 60.9854 | 118.4300 | 40.9854 | 89.29134 | 2 | 2 |
| Q5 | 0.3330 | 0.3321 | 67.8701 | 139.4970 | 57.8701 | 130.321 | 2 | 2 |
| Q6 | 0.8003 | 0.9250 | 73.8712 | 140.2586 | 100.8712 | 200.4531 | 3 | 3 |
| Q7 | 1.1002 | 0.9800 | 73.8542 | 151.3496 | 179.4142 | 435.6524 | 3 | 3 |
| Q8 | 5.9000 | 5.880 | 74.8710 | 150.2396 | 201.7710 | 530.2351 | 3 | 3 |
| Q9 | 7.8341 | 7.997 | 80.2010 | 158.4399 | 211.0110 | 542.4763 | 5 | 8 |
| Q10 | 30.213 | 32.345 | 121.3091 | 231.4954 | 250.3091 | 1053.0951 | 5 | 11 |
| Q11 | 38.967 | 36.867 | 156.0004 | 299.1496 | 230.1004 | 1103.4201 | 6 | 13 |
| Q12 | 55.343 | 57.687 | 200.1044 | 421.0394 | 250.1234 | 1732.0410 | 8 | 19 |
| Q13 | 55.823 | 60.347 | 311.4103 | 1145.9345 | 511.2203 | 3390.1306 | 10 | 25 |
| Q14 | 59.433 | 98.341 | 337.2134 | 3512.04501 | 637.2004 | 10458.9410 | 10 | 38 |

Table 2: queries with response time
with different data size and the number of joins

These set of fourteen queries starts with simple query and gradually are complex; we used the three types of datasets explained in table 1. The query performance time is evaluated for both our RMTT approach and RDF-3X approach. From table 2, it's clear that RMTT is faster than RDF-3X. For DBLP dataset RMTT performs a factor of 1-2 faster than RDF-3X, and for Dpedia performs 3-4 faster than RDF-3X approach. For CAS-Ibri dataset the queries response time are almost the same, because the characteristics of this type of data has no more self-joins.

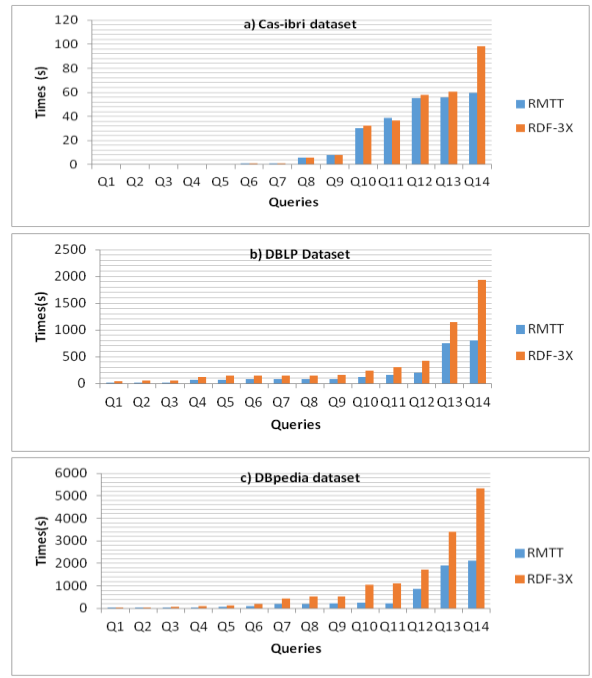

Figure 5: response times for a) CAS-ibri
b) DBLP  c) DBpedia

For scalability issue, we evaluate the number of self-joins (if its) using the DBpedia dataset against RMTT and RDF-3X approaches. Table 2 and figure 5 shows that for the first seven queries there is no difference between the two approaches. From query 8 to query 10 the difference in the number of self-joins between the two approaches are very small. From query 11 until query 14 which is more complex our proposed RMTT approach reduced the number of self-joins more than 30% compared with RDF-3X approach. This means the performance of our improves with more complex queries and the size of the data.

## 5. Conclusion and Future Work

In this paper, we have proposed the RMTT approach that comes as an alternative solution for the current approaches triple table, property table approach and vertically partitioned approach. Because they still suffer from weak data locality which affects the storage and query performance costs. RMTT approach is just a partitioning step for making high performance and efficient management of RDF data. In the future work, we try to demonstrate a good indexing mechanism which is suitable to RMTT approach. That works together with partitioning step and query the evaluation to achieve efficient and scalable management of RDF store.

**Appendix**

Query:1

PREFIX rdf: <http://www.w3.org/1999/02/22-rdf-syntax-ns#>
PREFIX ub: <http://www.lehigh.edu/~zhp2/2004/0401/univ-bench.owl#>
SELECT ?X
WHERE
{?X rdf:type ub:GraduateStudent .
?X ub:takesCourse
http://www.Department0.University0.edu/GraduateCourse0}

Query:2

PREFIX rdf: <http://www.w3.org/1999/02/22-rdf-syntax-ns#>
PREFIX ub: <http://www.lehigh.edu/~zhp2/2004/0401/univ-bench.owl#>
SELECT ?X, ?Y, ?Z
WHERE
{?X rdf:type ub:GraduateStudent .
?Y rdf:type ub:University .
?Z rdf:type ub:Department .
?X ub:memberOf ?Z .
?Z ub:subOrganizationOf ?Y .
?X ub:undergraduateDegreeFrom ?Y}

Query:3

PREFIX rdf: <http://www.w3.org/1999/02/22-rdf-syntax-ns#>
PREFIX ub: <http://www.lehigh.edu/~zhp2/2004/0401/univ-bench.owl#>
SELECT ?X
WHERE
{?X rdf:type ub:Publication .
?X ub:publicationAuthor
http://www.Department0.University0.edu/AssistantProfessor0}

Query:4

PREFIX rdf: <http://www.w3.org/1999/02/22-rdf-syntax-ns#>
PREFIX ub: <http://www.lehigh.edu/~zhp2/2004/0401/univ-bench.owl#>
SELECT ?X, ?Y1, ?Y2, ?Y3
WHERE
{?X rdf:type ub:Professor .
?X ub:worksFor <http://www.Department0.University0.edu> .
?X ub:name ?Y1 .
?X ub:emailAddress ?Y2 .
?X ub:telephone ?Y3}

Query:5

PREFIX rdf: <http://www.w3.org/1999/02/22-rdf-syntax-ns#>
PREFIX ub: <http://www.lehigh.edu/~zhp2/2004/0401/univ-bench.owl#>
SELECT ?X
WHERE
{?X rdf:type ub:Person .
?X ub:memberOf <http://www.Department0.University0.edu>}

Query:6

PREFIX rdf: <http://www.w3.org/1999/02/22-rdf-syntax-ns#>
PREFIX ub: <http://www.lehigh.edu/~zhp2/2004/0401/univ-bench.owl#>
SELECT ?X WHERE {?X rdf:type ub:Student}

Query:7

PREFIX rdf: <http://www.w3.org/1999/02/22-rdf-syntax-ns#>
PREFIX ub: <http://www.lehigh.edu/~zhp2/2004/0401/univ-bench.owl#>
SELECT ?X, ?Y
WHERE
{?X rdf:type ub:Student .
?Y rdf:type ub:Course .
?X ub:takesCourse ?Y .
<http://www.Department0.University0.edu/AssociateProfessor0>, ub:teacherOf, ?Y}

Query:8

PREFIX rdf: <http://www.w3.org/1999/02/22-rdf-syntax-ns#>
PREFIX ub: <http://www.lehigh.edu/~zhp2/2004/0401/univ-bench.owl#>
SELECT ?X, ?Y, ?Z
WHERE

{?X rdf:type ub:Student .
?Y rdf:type ub:Department .
?X ub:memberOf ?Y .
?Y ub:subOrganizationOf <http://www.University0.edu> .
?X ub:emailAddress ?Z}

Query:9

PREFIX rdf: <http://www.w3.org/1999/02/22-rdf-syntax-ns#>
PREFIX ub: <http://www.lehigh.edu/~zhp2/2004/0401/univ-bench.owl#>
SELECT ?X, ?Y, ?Z
WHERE
{?X rdf:type ub:Student .
?Y rdf:type ub:Faculty .
?Z rdf:type ub:Course .
?X ub:advisor ?Y .
?Y ub:teacherOf ?Z .
?X ub:takesCourse ?Z}

Query:10

PREFIX rdf: <http://www.w3.org/1999/02/22-rdf-syntax-ns#>
PREFIX ub: <http://www.lehigh.edu/~zhp2/2004/0401/univ-bench.owl#>
SELECT ?X
WHERE
{?X rdf:type ub:Student .
?X ub:takesCourse <http://www.Department0.University0.edu/GraduateCourse0>}

Query:11

PREFIX rdf: <http://www.w3.org/1999/02/22-rdf-syntax-ns#>
PREFIX ub: <http://www.lehigh.edu/~zhp2/2004/0401/univ-bench.owl#>
SELECT ?X
WHERE
{?X rdf:type ub:ResearchGroup .
?X ub:subOrganizationOf <http://www.University0.edu>}

Query:12

PREFIX rdf: <http://www.w3.org/1999/02/22-rdf-syntax-ns#>
PREFIX ub: <http://www.lehigh.edu/~zhp2/2004/0401/univ-bench.owl#>
SELECT ?X, ?Y
WHERE
{?X rdf:type ub:Chair .
?Y rdf:type ub:Department .
?X ub:worksFor ?Y .
?Y ub:subOrganizationOf <http://www.University0.edu>}

Query:13

PREFIX rdf: <http://www.w3.org/1999/02/22-rdf-syntax-ns#>
PREFIX ub: <http://www.lehigh.edu/~zhp2/2004/0401/univ-bench.owl#>
SELECT ?X
WHERE
{?X rdf:type ub:Person .
<http://www.University0.edu> ub:hasAlumnus ?X}

Query:14

PREFIX rdf: <http://www.w3.org/1999/02/22-rdf-syntax-ns#>
PREFIX ub: <http://www.lehigh.edu/~zhp2/2004/0401/univ-bench.owl#>
SELECT ?X
WHERE {?X rdf:type ub:UndergraduateStudent}

## Acknowledgments

This work is founded by TRC (The Research Council) Sultanate of Oman from 2012 to 2015.